\documentclass[prb,preprint]{revtex4-1}
\usepackage{amsmath}
\usepackage{mathrsfs}
\usepackage{amsfonts}
\usepackage{graphicx}

\begin{document}
\title{Typical length scales in conducting disorderless networks}

\author{M. Mart\'inez-Mares}
\email{moi@xanum.uam.mx}
\affiliation{Departamento de F\'isica, Universidad Aut\'onoma 
Metropolitana-Iztapalapa, Apartado Postal 55-534, 09340 Ciudad de M\'exico, 
Mexico}

\author{V. Dom\'inguez-Rocha}
\email{vdr@fis.unam.mx, vidomr@gmail.com}
\affiliation{Instituto de Ciencias F\'isicas, Universidad Nacional 
Aut\'onoma de M\'exico, Apartado Postal 48-3, 62210 Cuernavaca, Mor., 
Mexico}

\author{A. Robledo}
\email{robledo@fisica.unam.mx}
\affiliation{Instituto de F\'isica y Centro de Ciencias de la Complejidad,
Universidad Nacional Aut\'onoma de M\'exico, Apartado Postal 20-364, 01000
Ciudad de M\'exico, Mexico}

\begin{abstract}
We take advantage of a recently established equivalence, between the intermittent
dynamics of a deterministic nonlinear map and the scattering matrix properties of a 
disorderless double Cayley tree lattice of connectivity $K$, to obtain general electronic
transport expressions and expand our knowledge of the scattering properties at the
mobility edge. From this we provide a physical interpretation of the generalized localization length.
\end{abstract}
\maketitle
\section{Introduction}
\label{intro}

Very recently, it has been found that the electronic scattering properties of a layered linear
periodic structure and those of a regular nonlinear network model are described exactly by 
the dynamics of intermittent low-dimensional nonlinear maps~\cite{MM-R,Jiang,Victor2013}. 
The presence of these maps is a consequence of the combination rule of scattering matrices
when the scattering systems are built via consecutive replication of an element or motif. This
new insight implies an  equivalence between wave transport phenomena in classical wave
systems, or  electronic transport through quantum systems, and the dynamical properties of
low-dimensional nonlinear maps, specially at the onset of chaos. This is a remarkable property
in that a system with many degrees of freedom experiences a radical reduction of these,
so that its description is completely provided by only a few variables.

In particular, the band structure associated with scatterers arranged as a regular double Cayley
tree (see Fig.1) corresponds to dynamical properties of attractors of dissipative low-dimensional
nonlinear maps~\cite{MM-R}. The properties of the dimensionless conductance in the crystalline
limit reflect the periodic or chaotic nature of the attractors. The  transition between the insulating
to conducting phases can be seen as the transition along one of the known routes to (or out of)
chaos, the tangent bifurcation that exhibits intermittent dynamics in its vicinity~\cite{Schuster}.
While the conductance displays an exponential decay with size in the evolution towards the
crystalline limit, it obeys instead a $q$-exponential form at the transition. A similar behavior can
be found for a locally periodic structure where the wave function also decays exponentially in a
regular band (or regular attractor) and a $q$-exponential decay with system size at the mobility
edge (or onset of chaos)~\cite{Victor2013}. In the latter model, the typical decay length is
related to the mean free path. It is expected that the same occurs for the 
former model. 

In the present paper we generalize our treatment for the scattering properties 
across a double Cayley tree of arbitrary connectivity $K$. We focus on the 
conductance. Our purpose is to find the most general expressions for the band 
edges at the borderline behavior of the conductance toward the crystalline limit.
We also determine a general expression for the length scale at the transition.
In the next section we establish the recurrence relation for the scattering 
matrices of double Cayley trees of consecutive sizes. Next, we diagonalize
this relation in order to reduce the matrix expression to two equivalent nonlinear
maps for their eigenphases. This allows us to analize the system size dependence
of the sensitivity to initial conditions for the different types of attractors of the map,
including that at the transition. In Section~\ref{sect:transport} we consider the
implications for electronic transport. We conclude in Section~\ref{sec:6}.

\section{Scattering and deterministic maps}
\label{sec:sec1}

\subsection{Recurrence relation for the scattering matrix of a double Cayley 
tree}

\begin{figure}
\centering
\includegraphics[width=\columnwidth]{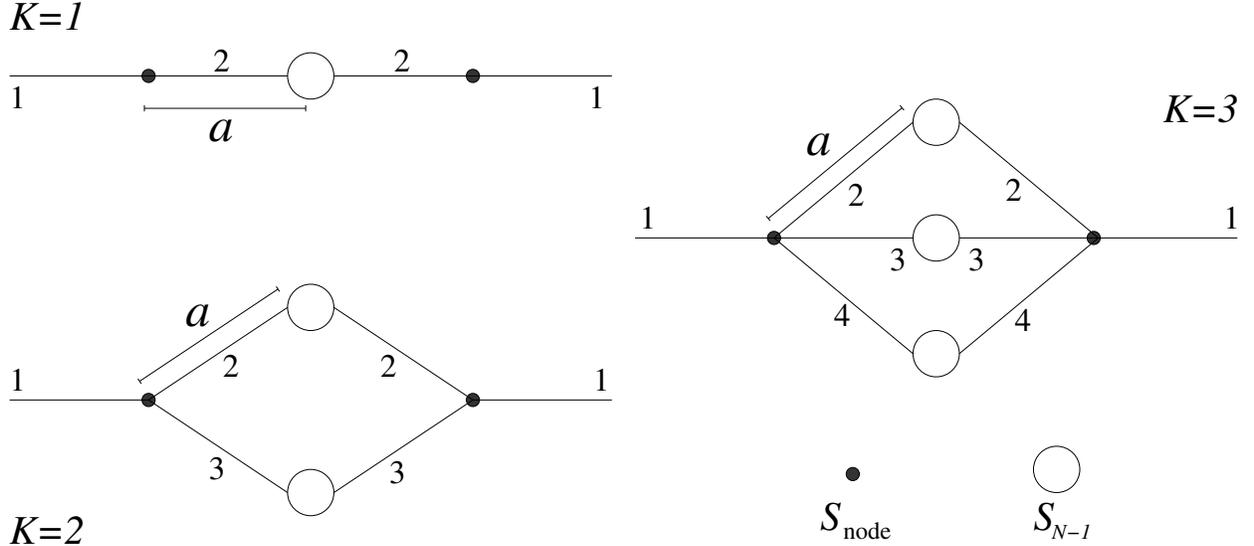}
\caption{Scheme of double Cayley trees of connectivity $K=1$, $K=2$ and $K=3$ 
at generation $N$. The leads indexed by 1 are coupled by the nodes, each 
described by the scattering matrix $S_{\rm{node}}$. $S_{N-1}$ is the scattering 
matrix at generation $N-1$.}
\label{fig:1}
\end{figure}

We consider an ordered double Cayley tree of connectivity $K\ge 1$. In 
Fig.~\ref{fig:1} we show the double Cayley trees for $K=1$, $K=2$ and $K=3$. We 
assume that the leads which connect two adjacent nodes, separated by a lattice 
constant $a$ (they are indexed by 2, 3, $\ldots K+1$ in the figure), are 
one-dimensional perfect wires. Also, we assume that each node is described by 
the same symmetric scattering matrix, which is of dimension $K+1$ and of the 
form
\begin{equation}
\label{eq:Snode}
S_{\rm{node}}=\left(
\begin{array}{c|ccccc}
S_{11} & S_{12} & S_{12} & \cdots & S_{12} & S_{12} \\ \hline
S_{12} & S_{22} & S_{23} & \cdots & S_{23} & S_{23} \\ 
S_{12} & S_{23} & S_{22} & \cdots & S_{23} & S_{23} \\
\vdots & \vdots & \vdots & \ddots & \vdots & \vdots \\
S_{12} & S_{23} & S_{23} & \cdots & S_{22} & S_{23} \\
S_{12} & S_{23} & S_{23} & \cdots & S_{23} & S_{22} \\
\end{array}
\right) . 
\end{equation}
The matrix $S_{\rm{node}}$ couples symmetrically the incoming lead 1 to the leads 2, 3, 
$\ldots K+1$, which are assumed to be equivalent. Flux conservation restricts 
$S_{\rm{node}}$ to be a unitary matrix; this condition is expressed by the three 
equations 
\begin{eqnarray}
\label{eq:unitary1}
|S_{11}|^2 + K|S_{12}|^2 & = & 1, \\
\label{eq:unitary2}
S^*_{11}S_{12} + S^*_{12} 
\left[ S_{22} + (K-1)S_{23} \right] & = & 0, \\ 
\label{eq:unitary3}
|S_{12}|^2 + |S_{22}|^2 + (K-1)|S_{23}|^2 & = & 1 .
\end{eqnarray}
Eq. (\ref{eq:unitary1}) restricts $S_{12}$ to 
$|S_{12}|\leq 1/\sqrt{K}$. Also, this equation can be 
written in terms of the reflection and transmission coefficients of the node, 
$R_{\rm{node}}$ and $T_{\rm{node}}$, respectively, as
\begin{equation}
R_{\rm{node}} + T_{\rm{node}} = 1, \quad\mbox{with}\quad 
R_{\rm{node}} = |S_{11}|^2 \quad\mbox{and}\quad
T_{\rm{node}} = K|S_{12}|^2.
\end{equation}

A recursion relation for the scattering matrix can be found using the 
combination rule of scattering matrices. We obtain the $2\times 2$ scattering 
matrix at a given generation $N$, $S_N$, by coupling $K$ scattering matrices 
$S_{N-1}$, at the previous generation, by means of the $K+1$-dimensional 
scattering matrices of the nodes. The result is  
\begin{equation}
\label{eq:STOT}
S_N=S_{PP} + 
S_{PQ} 
\frac{1}{I_{2K}-\mathrm{e}^{2\mathrm{i}ka}\mathbf{S}_{N-1} S_{QQ}} 
\mathrm{e}^{2\mathrm{i}ka}\mathbf{S}_{N-1} S_{QP},
\end{equation}
where $I_n$ denotes the $n\times n$ identity matrix and $\mathbf{S}_{N-1}$ is 
the $2K\times 2K$ matrix 
\begin{equation}
\label{eq:SNminusone}
\mathbf{S}_{N-1} = \left(
\begin{array}{cccc}
S_{N-1} & 0_2 & \cdots & 0_2 \\ 
0_2 & S_{N-1} & \cdots & 0_2 \\
\vdots & \vdots & \ddots & \vdots \\
0_2 & 0_2 & \cdots & S_{N-1} 
\end{array}
\right),
\end{equation}
where $0_n$ denotes the $n\times n$ null matrix. Here, $S_{PP}=S_{11}I_2$, it 
is a $2\times 2$ matrix that gives the reflection to the outside of the 
system, while $S_{QQ}$ is a $2K\times 2K$ matrix, responsible for the multiple 
scattering inside the system, which is given by 
\begin{equation}
\label{eq:PPQQ}
S_{QQ}=\left(
\begin{array}{cccc}
S_{22}I_2 & S_{23}I_2 & \cdots & S_{23}I_2 \\
S_{23}I_2 & S_{22}I_2 & \cdots & S_{23}I_2 \\
\vdots & \vdots & \ddots & \vdots \\
S_{23}I_2 & S_{23}I_2 & \cdots & S_{22}I_2
\end{array}
\right);
\end{equation}
$S_{PQ}$ and $S_{QP}$ gives the transmission from the outside to inside of the 
system and viceversa, respectively. They are the $2\times 2K$ and $2K\times 2$ 
matrices 
\begin{equation}
S_{PQ} = S_{12}\left(
\begin{array}{cc}
I_2 & I_2 
\end{array}
\right)
\quad\mbox{and}\quad
S_{QP} = S_{12} \left(
\begin{array}{c}
I_2 \\ I_2 
\end{array}
\right) ,
\end{equation}
respectively. Therefore, Eq.~(\ref{eq:STOT}) is simplified to the expression
\begin{equation}
\label{eq:recursion}
S_N = S_{11}I_2 + 
\frac{K \left( S_{12}\mathrm{e}^{\mathrm{i}ka}\right)^2}
{I_2 - \mathrm{e}^{2\mathrm{i}ka}[S_{22}+(K-1)S_{23}]S_{N-1}}
S_{N-1}, 
\end{equation}
whose physical interpretation is clear and the same as in Eq.~(\ref{eq:STOT}). 
The factor $K$ in front of the second term on the right hand side of 
Eq.~(\ref{eq:recursion}) is due to the identical couplings of lead 1 to leads 2, 
3, $\ldots K$. 

\subsection{Reduction to nonlinear iterated maps}

The structure of the matrix $S_{\rm{node}}$ in Eq.~(\ref{eq:Snode}) leads us to 
a left-right symmetric system in the presence of time reversal invariance. 
In that case, $S_N$ has a block symmetric structure, namely 
\begin{equation}
\label{eq:Symmetric}
S_N = \left( 
\begin{array}{cc}
r_N & t_N \\ t_N & r_N
\end{array}
\right), 
\end{equation}
which is easily diagonalized by a $\pi/4$-rotation, 
\begin{equation}
\left(\begin{array}{cc}
\textrm{e}^{\textrm{i}\theta_N} & 0\\
0 & \textrm{e}^{\textrm{i}\theta'_N}
\end{array}\right)=
R_0 S_N R_0^T,
\quad
R_0 =\frac{1}{\sqrt{2}}\left(
\begin{array}{cc}
1 & 1\\
-1 & 1
\end{array}
\right),
\label{eq:diagonal}
\end{equation}
with $R_0^T$ the transpose of $R_0$. Here, the eigenphases $\theta_N$ and 
$\theta'_N$ are given by $\textrm{e}^{\textrm{i}\theta_N}=r_N+t_N$ and 
$\textrm{e}^{\textrm{i}\theta'_N}=r_N-t_N$. The diagonal form of the recursion 
relation (\ref{eq:recursion}) lead to a nonlinear map satisfied by both 
eigenphases. For instance, $\theta_N=f(\theta_{N-1})$, where 
\begin{equation}
f(\theta_{N-1}) = -\theta_{N-1} + 2\arctan 
\frac{\mathrm{Im} \left( S_{11}S^*_{12}\mathrm{e}^{-\mathrm{i}ka} + 
S_{12} \mathrm{e}^{\mathrm{i}ka} 
\mathrm{e}^{\mathrm{i}\theta_{N-1}}\right)}
{\mathrm{Re} \left( S_{11}S^*_{12}\mathrm{e}^{-\mathrm{i}ka} + S_{12} 
\mathrm{e}^{\mathrm{i}ka} 
\mathrm{e}^{\mathrm{i}\theta_{N-1}}\right)}.
\label{eq:map}
\end{equation}
Here, we used the unitarity condition of $S_{\rm{node}}$ through 
Eqs.~(\ref{eq:unitary1}) and (\ref{eq:unitary2}). We note that this map depends 
on $S_{12}$ and on the phase of $S_{11}$ through Eq.~(\ref{eq:unitary2}). 
The dependence on $K$ is implicit through Eq.~(\ref{eq:unitary1}). If we assume 
that at the two branches of the double Cayley tree are perfectly joined at the 
middle. The initial conditions for both maps at $N=0$ are $\theta_0=0$ and 
$\theta'_0=\pi$. 

\begin{figure}
\centering
\includegraphics[width=\columnwidth]{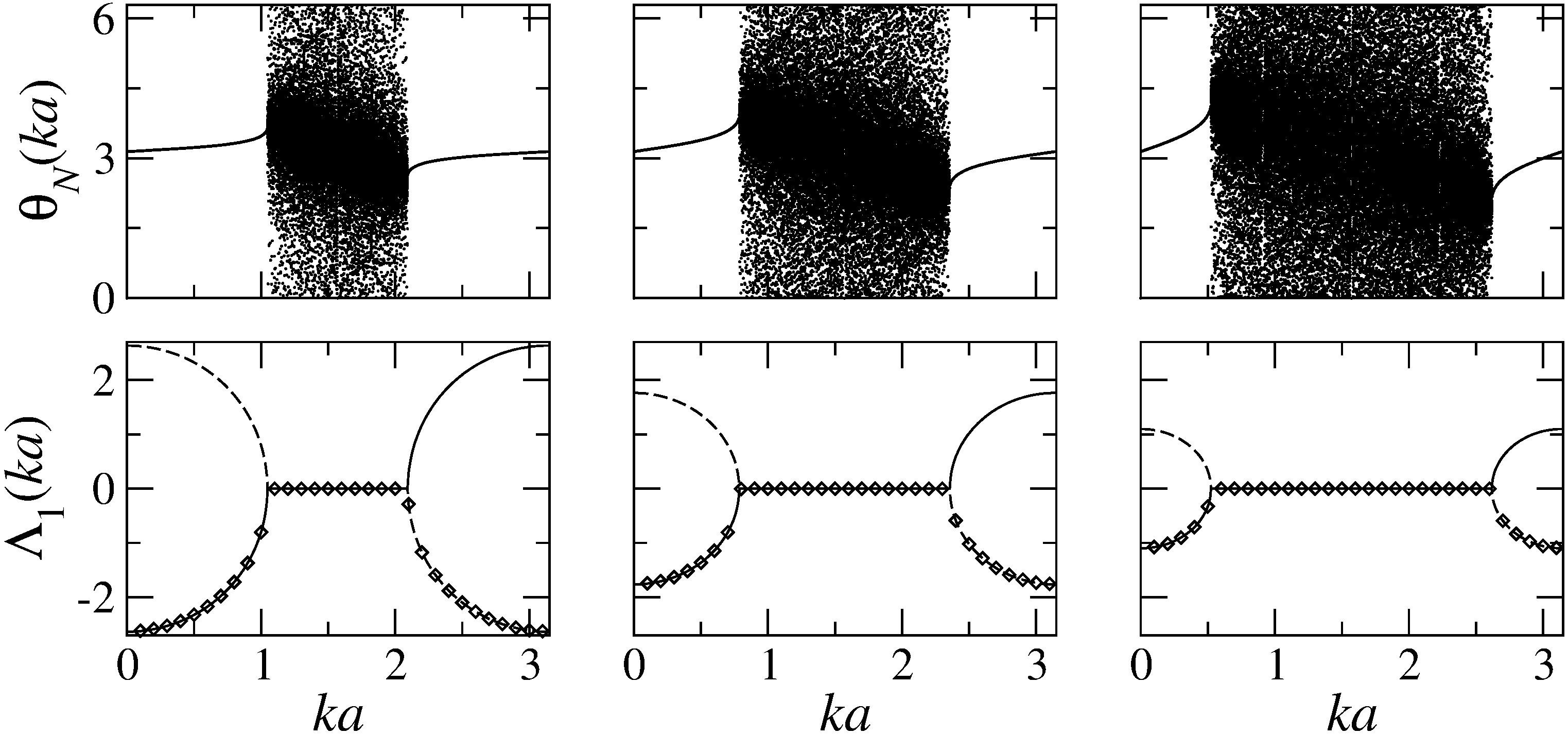} 
\caption{Upper panels: Bifurcation diagrams for a double Cayley tree of 
connectivity $K=1$ (left), $K=2$ (middle) and $K=3$ (right). For each value of 
$ka$ we plot only the last 50 iterations of $N=1000$ starting with an initial 
condition $\theta_0=0$. Lower panels: Finite $N$ Lyapunov coefficient. The 
diamonds correspond to the iteration $N=1000$ of Eq.~(\ref{eq:XiN}), with an 
initial condition $\theta_0=0$. Continuous and dashed lines represent the 
Lyapunov coefficient (\ref{eq:Lyapunovexact}) for the two roots of 
Eq.~(\ref{eq:fixed}). For all cases $S_{12}=1/2$ and $S_{11}=-\sqrt{1-K/4}$. }
\label{fig:map}
\end{figure}

The bifurcation diagrams corresponding to $K=1,\,2$ and 3 are shown in the 
upper panels of Fig.~\ref{fig:map} for $S_{11}=-\sqrt{1-K|S_{12}|^2}$ with 
$S_{12}=1/2$ and an initial condition $\theta_0=0$. We can observe that the map 
(\ref{eq:map}) presents ergodic windows (we show only one on each panel) between 
windows of periodicity 1. This figure suggests that $\theta_N$ reaches a fixed 
point solution. Looking for those fixed point solutions of Eq.~(\ref{eq:map}), 
we find that
\begin{equation}
\label{eq:fixed}
\mathrm{e}^{\mathrm{i}\theta_{\infty}(ka)} = 
\left\{\begin{array}{ccc}
\textrm{e}^{\textrm{i}\theta_{\pm}(ka)} & \mbox{for} & |\mathrm{Re} 
( S_{12}\mathrm{e}^{\mathrm{i}ka} )| > \sqrt{K} |S_{12}|^2 \\
w_{\pm}(ka) & \mbox{for} & |\mathrm{Re} 
( S_{12}\mathrm{e}^{\mathrm{i}ka} ) | \leq \sqrt{K} |S_{12}|^2 
\end{array}
\right. ,
\end{equation}
where 
\begin{eqnarray}
\label{eq:infty}
\mathrm{e}^{\mathrm{i} \theta_{\pm}(ka)} & = & \frac{ \pm 
\sqrt{ \left[ \mathrm{Re} \left( S_{12} \mathrm{e}^{\mathrm{i}ka} 
\right)\right]^2 -K|S_{12}|^4} + \mathrm{i}\, \mathrm{Im}\left( 
S_{12}\mathrm{e}^{\mathrm{i}ka} \right) } 
{S^*_{11}S_{12}\mathrm{e}^{\mathrm{i}ka}}, \\
w_{\pm}(ka) & = & -\frac{ \pm 
\sqrt{ \left[ \mathrm{Re} \left( S_{12} \mathrm{e}^{\mathrm{i}ka} 
\right)\right]^2 -K|S_{12}|^4} + \mathrm{Im}\left( 
S_{12}\mathrm{e}^{\mathrm{i}ka} \right) } 
{\mathrm{i}\,S^*_{11}S_{12}\mathrm{e}^{\mathrm{i}ka}}.
\end{eqnarray}
Note that for each value of $ka$ there are two solutions. In the windows of 
periodicity 1, one of these solutions coincide with the result from the 
iteration shown in Fig.~\ref{fig:map}. This limiting value $\theta_{\infty}(ka)$ 
of $\theta_N(ka)$ corresponds to an attractor. The second solution that does 
not appear in Fig.~{\ref{fig:map}} corresponds to a repulsor. If an initial 
condition $\theta_0$ is just the value of the repulsor, that is 
$\theta_0=\theta_{\rm{rep}}$, the solution will remain there forever. Any other 
initial condition will converge to the attractor. In these windows the fixed 
point solutions are of the form $\mathrm{e}^{\mathrm{i}\theta_\pm(ka)}$, as 
expected for an eigenphase. However, in an ergodic window the fixed point 
solutions of Eq.~(\ref{eq:map}) do not have modulus 1, but they are of the form 
$w_{\pm}(ka)=|w_{\pm}(ka)|\textrm{e}^{\mathrm{i}\theta(ka)}$, with $\theta(ka)$ 
being the value around which $\theta_N(ka)$ fluctuates with an invariant 
density. These solutions are marginally stable~\cite{Wolf}. From 
Eq.~(\ref{eq:fixed}) we see that the critical values $k_c$ of $k$ that separates 
the ergodic and periodic windows, critical attractors, satisfy that
\begin{equation}
\left| \mathrm{Re}\left( S_{12} \mathrm{e}^{\mathrm{i}k_ca} \right) 
\right| = \sqrt{K} \left| S_{12}\right|^2.
\end{equation}
At these critical attractors, $\theta(k_ca)\equiv\theta_c$, where $\theta_c$ 
are the points of tangency given by 
\begin{equation}
\tan{\theta_c} = \frac{\mathrm{Im} \left(\mathrm{i}\,S_{11}S^*_{12} 
\mathrm{e}^{-\mathrm{i}k_ca} \right)}
{\mathrm{Re} \left(\mathrm{i}\,S_{11}S^*_{12} 
\mathrm{e}^{-\mathrm{i}k_ca} \right)}.
\end{equation}

\subsection{Sensitivity to initial conditions}

The dynamics of the map (\ref{eq:map}) is characterized by the sensitivity to 
initial conditions. For finite $N$, it is defined by~\cite{MM-R}
\begin{equation}
\Xi_N\equiv \mathrm{e}^{N\Lambda_1(N)} \equiv 
\left| \frac{\mathrm{d}\theta_N}{\mathrm{d}\theta_0} \right|,
\end{equation}
where $\theta_0$ is the initial condition and $\Lambda_1(N)$ is the finite $N$ 
Lyapunov exponent. From Eq.~(\ref{eq:map}), we find the following recursive 
relation for $\Xi_N$,
\begin{equation}
\label{eq:XiN}
\Xi_N(ka) = 
\frac{K|S_{12}|^4}
{\left| S^*_{12}\mathrm{e}^{-\mathrm{i}ka} + 
S^*_{11}S_{12} \mathrm{e}^{\mathrm{i}ka}
\mathrm{e}^{\mathrm{i}\theta_{N-1}(ka)} \right|^2} \, 
\Xi_{N-1}(ka).
\end{equation}
In the limit $N\to\infty$, $\Xi_N\to\xi_N$ with $\xi_N$ being the sensitivity 
to initial conditions, defined by
\begin{equation}
\xi_N(ka) = \mathrm{e}^{N\lambda_1(ka)}, N\gg 1,
\label{eq:xi}
\end{equation}
with $\lambda_1(ka)$ the Lyapunov exponent 
\begin{equation}
\label{eq:Lyapunovexact}
\lambda_1(ka) = \ln 
\frac{K|S_{12}|^4}
{\left| S^*_{12}\mathrm{e}^{-\mathrm{i}ka} + 
S^*_{11}S_{12} \mathrm{e}^{\mathrm{i}ka}
\mathrm{e}^{\mathrm{i}\theta_{\infty}(ka)} \right|^2}.
\end{equation}
In the lower panels of Fig.~\ref{fig:map} we show the behavior of 
$\lambda_1(ka)$ (or $\Lambda_1(ka)$ in the limit $N\gg 1$) for the three cases: 
$K=1$, 2 and 3. We observe that in the windows of periodicity 1 the theoretical 
result $\lambda_1(ka)$ of Eq.~(\ref{eq:Lyapunovexact}) shows two values for a 
given $ka$, which correspond to both roots expressed in Eq.~(\ref{eq:fixed}). 
For the repulsor, $\lambda_1(ka)>0$ indicating that $\xi_N(ka)$ diverges 
exponentially, while at the attractor $\lambda_1(ka)<0$ and $\xi_N(ka)$ decays 
exponentially with a typical length scale given by 
\begin{equation}
\label{eq:zeta1}
\zeta_1(ka) = \frac{a}{|\lambda_1(ka)|} .
\end{equation}
As it happens for the fixed-point solutions, only the solution for the attractor 
agrees with the $\Lambda_1(ka)$ obtained from the iteration of Eq.~(\ref{eq:XiN}).
For the ergodic windows we have $\lambda_1(ka)=0$, so nothing we can say 
about the $N$-dependence of $\xi_N(ka)$. However, in those windows $\Xi_N(ka)$ 
oscillates (not shown here) with $N$.

\begin{figure}
\includegraphics[width=\columnwidth]{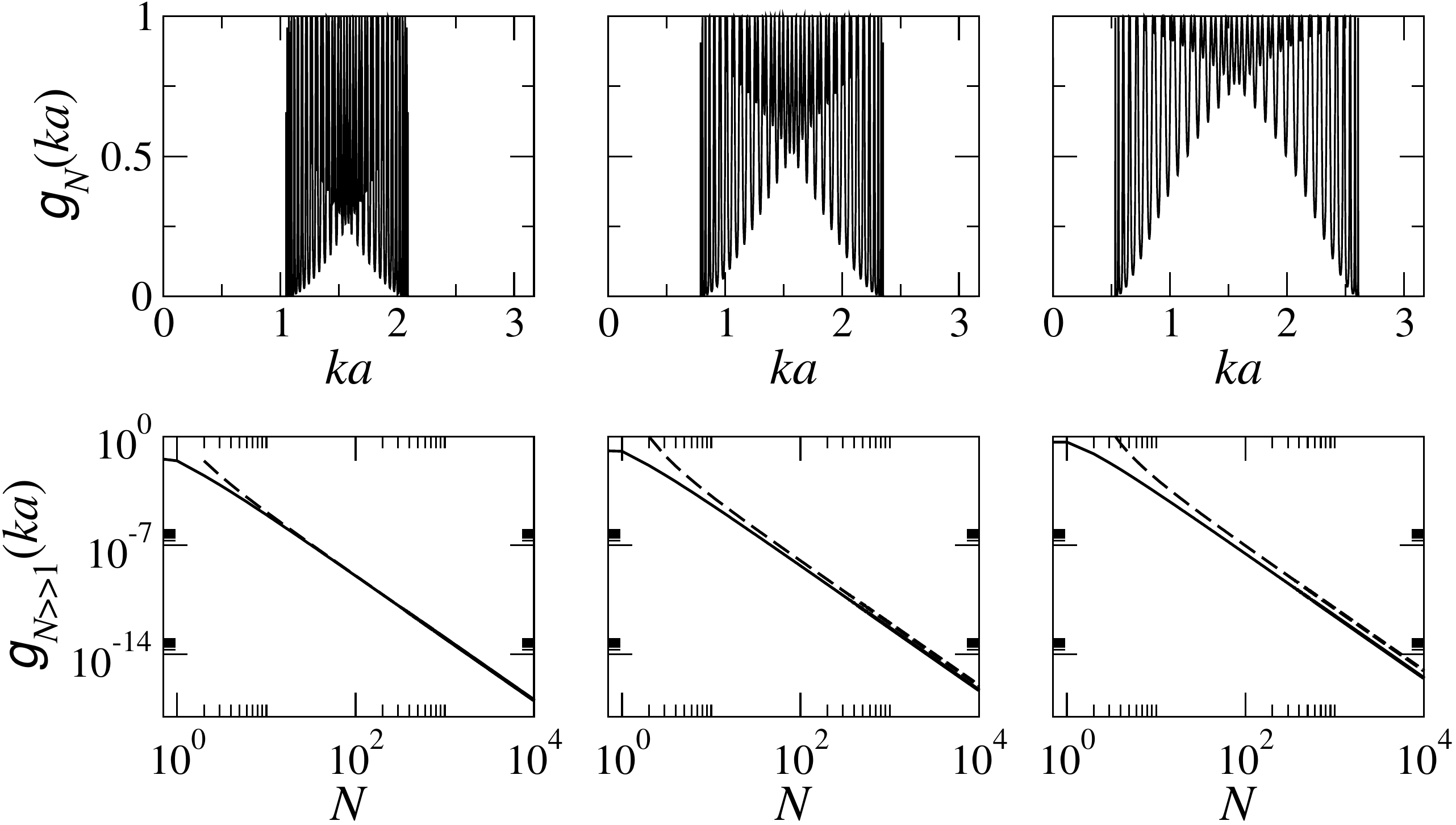}
\caption{Upper panels: Bands of the dimensionless conductance for a double 
Cayley tree of connectivity $K=1$ (left), $K=2$ (middle), and $K=3$ (right), for 
$N=30$ and initial conditions $\theta_0=0$ and $\theta'_0=\pi$. Lower panels: 
Dimensionless conductance at the critical attractors for $N\gg 1$ at 
$k_ca=\pi/3$ for $K=1$, $k_ca=\pi/4$ for $K=2$, and $k_ca=\pi/6$ for $K=3$. 
Dashed lines represent the theoretical result of Eq.~(\ref{eq:gncritical}) with 
$S_{12}=1/2$ and $S_{11}=-\sqrt{1-K/4}$.}
\label{fig:lambda1}
\end{figure}

At the critical attractors (those for the tangent bifurcations) we find that
\begin{equation}
\theta_N-\theta_c = (\theta_{N-1}-\theta_c) \pm 
\sqrt{\frac{R_{\rm{node}}}{T_{\rm{node}}}}
\left| \theta_{N-1}-\theta_c \right|^2 + 
\cdots.
\end{equation}
This local nonlinearity leads, via the functional composition renormalization group
fixed-point map~\cite{Schuster}, to a $q$-exponential expression for the sensitivity
for any $N$, namely~\cite{Baldovin}
\begin{equation}
\xi_N =
\left(1- \frac{1}{2} \lambda_{3/2}N\right)^{-2},
\end{equation}
where $\lambda_{3/2}$ is the $q$-generalized Lyapunov coefficient for $q=3/2$, 
which is given by $\lambda_{3/2}=\pm 2\sqrt{R_{\rm{node}}/T_{\rm{node}}}$. The 
plus and minus signs corresponds to trajectories at the left and right of the 
point of tangency $\theta_c$. When $\theta_N-\theta_c>0$, $\xi_N$ grows with 
$N$ faster than exponential and when $\theta_{N-1}-\theta_c<0$, $\xi_N$ decays 
with $N$ with a power-law behaviour, the typical decay length being given by
\begin{equation}
\label{eq:zeta3/2}
\zeta_{3/2} = \frac{a}{|\lambda_{3/2}|} = 
\frac{a}{2} \sqrt{\frac{T_{\rm{node}}}{R_{\rm{node}}}}.
\end{equation}
This result on diverging duration of the laminar episodes of intermittency and 
large $N$ intervals of vanishing $\Xi_N$ between increasingly large spike 
oscillations.

\section{Consequences for the electronic transport}
\label{sect:transport}

According to the Landauer formula, the dimensionless conductance $g_{_N}$ at the 
generation $N$ (conductance $G_N$ in units of $2e^2/h$) is just the 
transmission coefficient $|t_N|^2$~\cite{Buttiker}. Using 
Eqs.~(\ref{eq:Symmetric}) and (\ref{eq:diagonal}) we find a recursion relation 
for the conductance, namely
\begin{equation}
g_{_N}(ka) = g_{_{N-1}}(ka) \, 
\frac{\Xi_{N}(ka)}{\Xi_{N-1}(ka)}\,
\frac{\Xi'_{N}(ka)}{\Xi'_{N-1}(ka)}.
\end{equation}
By iteration of this recursion relation we obtain
\begin{equation}
g_{_N}(ka) = \Xi_N(ka)\, \Xi'_N(ka).
\end{equation}

For the initial conditions $\theta_0=0$ and $\theta'_0=\pi$, 
$\Lambda_1\to\lambda_1$ and $\Lambda'_1\to\lambda_1$ for $N\gg 1$, therefore
\begin{equation}
g_{_N}(ka) = \mathrm{e}^{2N\lambda_1(ka)}.
\end{equation}
This means that in the ergodic windows, where $\lambda_1(ka)=0$, the conductance 
does not decay but oscillates with $N$. However, in the windows of 
periodicity 1 $\lambda_1(ka)<0$ and $g_{_N}(ka)$ shows an exponential decay with 
$N$, whose typical length scale is $\zeta_1(ka)$ of Eq.~(\ref{eq:zeta1}). In 
analogy to the scaling behaviour of the conductance with the size of a 
disordered system~\cite{Lee,Beenakker}, we name $\zeta_1(ka)$ localization 
length. In the upper panels of Fig.~\ref{fig:lambda1} we observe the bands of 
the dimensionless conductance, where the windows of periodicity 1 correspond 
to forbidden bands, while the chaotic windows relate to allowed bands. Critical 
attractors also correspond to the band edges, at which we expect that
\begin{equation}
\label{eq:gncritical}
g_{_N} = \xi_N^2 = 
\left(1- \frac{1}{2} \lambda_{3/2}N\right)^{-4}.
\end{equation}
That is, the conductance shows a power law decay, such that $\zeta_{3/2}$, 
in Eq.~(\ref{eq:zeta3/2}), is the typical decay length over large $N$ intervals 
located between increasingly large spike oscillations. In analogy with $\zeta_1$ 
we define $\zeta_{3/2}$ to be a localization length too. What is interesting about 
this length scale is its relation with the mean free path $\ell$ defined 
as~\cite{Mello} 
\begin{equation}
\frac{1}{\ell} = \frac{1}{a}\,\cdot\, \frac{R_{\rm{node}}}{1-R_{\rm{node}}}.
\end{equation}
This implies that the localization length at the critical attractors is one half 
of the geometric mean of the mean free path and the lattice constant,
\begin{equation}
\zeta_{3/2} = \frac{1}{2} \,
\sqrt{\ell\, a},
\end{equation}
which can be interpreted as the distance traveled by an electron before 
scattering. In the lower panels of Fig.~\ref{fig:lambda1} we observe that the 
power law decay fits very well with the scaling behavior with $N$ of the 
conductance. 

\section{Conclusions}
\label{sec:6}

We presented a generalized approach for the determination the dimensionless
conductance of a double Cayley tree of charge scatterers of arbitrary connectivity.
This is done by studying its scattering properties, as a function of the system size
(generation), through the sensitivity to initial conditions of the nonlinear map satisfied
by the eigenphases of the scattering matrix associated with the system. In the limit
of a very large system the conducting and insulating bands correspond, respectively,
to marginally chaotic windows and windows of periodicity 1. While in the conducting 
bands the conductance oscillates with the system size, in the insulating phase 
it displays an exponential decay with the system size, with a typical length scale, 
as in the scaling theory of localization. However, at the transition, on a band edge, 
when the conductance decays as a power law, the typical length scale is a
$q$-generalized localization length, which is the geometric mean of the mean free 
path and the lattice constant.

The insulator to conductor transition in electronic transport systems is a condensed-matter
phenomenon that still poses significant challenges before unabridged understanding is attained. The
occurrence of a robust analogy between the size-dependent properties of an idealized network
model of electron scatterers and the dynamical properties of a low-dimensional nonlinear map
displaying tangent bifurcations is a remarkable finding. On the one hand this connection makes
possible an exact determination of the conductance at the transition, while on the other hand 
reveals how a system composed of many degrees of freedom can undergo a drastic simplification.

\section{Acknowledgement}

V. Dom\'inguez-Rocha thanks DGAPA, UNAM, Mexico, for financial support. 
M. Mart\'inez-Mares is grateful to the Sistema Nacional de Investigadores, 
Mexico. A. Robledo acknowledges support from DGAPA-UNAM-IN103814
and CONACyT-CB-2011-167978 (Mexican Agencies).

\end{document}